# An unexplored valley of binary packing: The loose jamming state

Si Suo[1, 2, 3#], Chongpu Zhai[1#], Minglong Xu[1*], Marc Kamlah[2], Yixiang Gan[4,5**]

[1] *State Key Laboratory for Strength and Vibration of Mechanical Structures, School of Aerospace, Xi'an Jiaotong University, Xi'an, China*

[2] *Institute for Applied Materials, Karlsruhe Institute of Technology (KIT), Germany*

[3] *Department of Engineering Mechanics, FLOW Centre, KTH, Stockholm SE-100 44, Sweden*

[4] *School of Civil Engineering, The University of Sydney, Sydney, NSW 2006, Australia*

[5] *The University of Sydney Nano Institute (Sydney Nano), The University of Sydney, Sydney, NSW 2006, Australia*

We present a theoretical prediction on random close packing factor $\phi_{RCP}^b$ of binary granular packings based on the hard-sphere fluid theory. An unexplored regime is unravelled, where the packing fraction $\phi_{RCP}^b$ is smaller than that of the mono-sized one $\phi_{RCP}^m$, i.e., the so-called loose jamming state. This is against our common perception that binary packings should always reach a denser packing than mono-sized packings at the jamming state. Numerical evidence further supports this prediction and confirms the regime location in the size ratio and mole fraction ($R_r$-$X_s$) space, where the size ratio $R_r$ is close to 1, and the mole fraction of the smaller sphere $X_s$ close to 0. This extreme regime remains unreported in existing literature, yet significant for our fundamental understanding of binary packing systems.

\# Contribute equally.
\* Corresponding author. Email: mlxu@mail.xjtu.edu.cn.
\*\* Corresponding author. Email: yixiang.gan@sydney.edu.au

*Introduction*

The random close packing (RCP) problem has attracted extensive attention, which is of great importance not only in our exploration into the fundamental physics of glasses, liquids, colloidal systems [1], but also in facilitating optimisation of conveying, handling and processing different types of granular materials [2]. The RCP problem can be dated back to Bernal and Mason's experimental study on packings with identical spheres, in which the random close packing factor for mono-sized packing, $\phi_{RCP}^m$, was measured to be around 0.64 with an average coordination number, $Z_c^m$, of 6 [3]. Following experimental and numerical studies [4-8] reported that $\phi_{RCP}^m$ generally ranges from 0.61 to 0.69. Despite great progress in both experimental and numerical studies [9-12], we still lack a comprehensive understanding of RCP, especially for binary packings, in terms of the particle-scale driving mechanism, collective behaviour of jamming development [13, 14], tuneable properties including stiffness, mechanical stability [6], anti-crystallization [15, 16], thermal and electrical conduction [17, 18].

Few analytical approaches have been reported enabling satisfactory interpretation for definition, formation, characterization of mono-sized RCP. These models [19-22] tend to simplify the structure evolution of RCP by concentrating on the contacting neighbours of an individual particle, while ignoring influences of non-contacting surrounding particles. Very recently, Zaccone [23] proposed a simple route towards an analytical value of $\phi_{RCP}^m$ using a nearest-neighbour statistics as well as respecting the known closest packing ($Z_c^m = 12$, $\phi_{CP}^m \approx 0.74$). Noticeably, no extra fitting parameters are required in this scheme and the predicted $\phi_{RCP}^m$ highly matches the measurements from existed experiments and simulations.

Regarding binary granular packings, the mixture can usually fill more space than that of mono-sized packings, with the corresponding RCP factor $\phi_{RCP}^b > \phi_{RCP}^m$. Theoretical solutions for both mono-sized packings [24] and binary mixtures [25] were developed based on equilibrium statistical mechanics. The current focus is on situations with moderate or extremely small size ratio $R_r$ (small radius on large one) [10-12, 16, 26].

A quantity of empirical packing models were proposed based on the assumption that fine particles can fill the void among large particles [27-29]. However, the other near-boundary extreme region, i.e., $R_r \to 1$, has been seldom investigated, as an asymptotic trend to the mono-sized packing can be expected.

In this letter, we remedy to the unexplored region from theoretical and numerical aspects. Firstly, the Zaccone's scheme is extended to the binary packing problem, and at the near-boundary region the theoretical solution indicates an "abnormal" valley, where $\phi_{RCP}^b < \phi_{RCP}^m$, i.e., a so-called loose jamming state. Then, we further provide numerical evidence to support our theoretical prediction.

*Theory*

Here, the hard-sphere (HS) fluid theory is employed for estimating the partial radial distribution functions (RDF) in binary packing, allowing the statistical description of the local structure around a sphere. Specifically, three types of sphere-to-sphere interaction including small-small, large-large, and small-large pairs, with $g_{ss}(d)$, $g_{ll}(d)$ and $g_{sl}(d)$, respectively, indicate the probability density within a radial distance of $d$. Correspondingly, the contact value $Z_{c,ij}$ for the above three pair types ($i,j$ for $s$ or $l$) can be given by [23]

$$Z_{c,ij} = 4\pi\rho \int_0^{d_{ij}+\varepsilon} g_{ij}^0\, g_{ij}(d_{ij})\delta(r-d_{ij})r^2 dr, \qquad (1)$$

where $d_{ij}$ is the centre distance between two contacting spheres; $\rho$ is the number density, i.e. $\rho = (N_s + N_l)/V$ with $N_s$ and $N_l$ being the number of small and large spheres within a volume of $V$; $g_{ij}^0$ is a factor to be determined from the consistency conditions stated later. By weighting the contact value with mole fraction, the average coordination number $Z_c^b$ of a binary packing is

$$Z_c^b(X_s, R_r) = X_s(Z_{c,ss} + Z_{c,sl}) + X_l(Z_{c,ll} + Z_{c,ls}), \qquad (2)$$

where the mole fraction $X_i = N_i/(N_s + N_l)$ and the size ratio $R_r = d_s/d_l$. Naturally, the binary packing reduces to mono-sized packing, i.e., $Z_c^b(X_s, R_r)$ should be consistent with $Z_c^m$ under the following situations, i.e., the so-called consistency conditions:

$$Z_c^b(0, R_r) = Z_c^m; \ Z_c^b(1, R_r) = Z_c^m; \ Z_c^b(X_s, 1) = Z_c^m. \tag{3}$$

Moreover, substituting Eq. (1) into Eq. (2), $Z_c^b$ reads

$$Z_c^b = 4\pi\rho g^0[X_s d_s^3 \tilde{g}_{ss}^0 g_{ss}(d_{ss}) + X_l d_l^3 \tilde{g}_{ll}^0 g_{ll}(d_{ll}) + d_{sl}^3 \tilde{g}_{sl}^0 g_{sl}(d_{sl})], \tag{4}$$

where the factor $g_{ij}^0$ of Eq. (1) is expressed as $g_{ij}^0 = g^0 \tilde{g}_{ij}^0$. Here, $g^0$ is a normalization constant, and as followed by Zaccone's route [23], $g^0$ can be determined by introducing the conclusion of the mono-sized close packing ($Z_c^m = 12$, $\phi_{CP}^m \approx 0.74$); $\tilde{g}_{ij}^0$ is responsible for the consistency condition and therefore it is a function of $X_s$ and normalized within [0, 1]. Specifically, corresponding to the consistency conditions, the constraints on $\tilde{g}_{ij}^0$ are given respectively by

$$\tilde{g}_{ss}^0(0) = 0; \ \tilde{g}_{ss}^0(1) = 1; \tag{5}$$

$$\tilde{g}_{ll}^0(0) = 1; \ \tilde{g}_{ll}^0(1) = 0; \tag{6}$$

$$X_s \tilde{g}_{ss}^0(X_s) + (1 - X_s)\tilde{g}_{ll}^0(X_s) + \tilde{g}_{sl}^0(X_s) = 1. \tag{7}$$

Under the constraints of Eqs. (5)~(7), the possible format of $\tilde{g}_{ij}^0$ can be

$$\tilde{g}_{ss}^0(X_s) = X_s; \ \tilde{g}_{ll}^0(X_s) = 1 - X_s; \ \tilde{g}_{sl}^0(X_s) = 2X_s(1 - X_s). \tag{8}$$

Regarding $g_{ij}(d_{ij})$, the theory on additive HS mixtures, as an extension of the mono-component one, can provide a statistical solution, e.g., the BMCSL scheme [30] extended from the Carnahan-Starling equation [31]. It has been proved that the estimation on $g_{ij}(d_{ij})$ given by the original BMCSL expression is remarkably accurate for the moderate region of $X_s$ and $R_r$, while for regions $X_s \to 0$ or $X_s \to 1$ of interest in this work, deviation occurs [32, 33]. Thus, we employ a modified version which improved the BMCSL expression to fully satisfy the nine consistency conditions for binary mixtures by adding an additional term [33]. Moreover, in order to generate a quantitative description and better agree with the experimental and simulation results, two adjustable indexes $n_l$ and $n_s$ are introduced on the additional term whilst all consistency conditions in [33] are respected. The final expressions, where $g_{ij}^{\text{BMCSL}}(d_{ij})$ is the contact value of the RDF from the BMCSL expression, explicitly read

$$g_{ll}(d_{ll}) = g_{ll}^{\text{BMCSL}}(d_{ll}) + \left(\frac{X_s}{4}\frac{\xi_1\xi_2}{(1-\xi_3)^2}\frac{d_{ll}-d_{ss}}{d_{sl}}d_{ll}^2 d_{ss}\right)^{n_l}, \tag{9}$$

$$g_{ss}(d_{ss}) = g_{ss}^{\text{BMCSL}}(d_{ss}) + \left(\frac{X_l}{4}\frac{\xi_1\xi_2}{(1-\xi_3)^2}\frac{d_{ll}-d_{ss}}{d_{sl}}d_{ss}^2 d_{ll}\right)^{n_s}, \tag{10}$$

$$g_{sl}(d_{sl}) = g_{sl}^{\text{BMCSL}}(d_{sl}) + \left(\frac{1}{4}\frac{\xi_1\xi_2}{(1-\xi_3)^2}\frac{d_{ll}-d_{ss}}{d_{sl}}\frac{d_{ll}d_{ss}^3}{d_{sl}}\right), \text{ and} \tag{11}$$

$$\xi_n = \frac{\pi}{6}\rho \sum_i x_i d_{ii}^n. \tag{12}$$

In the original BMCSL scheme [24], both $n_l$ and $n_s$ equal 1. Alternative modifications based on the BMCSL scheme can be referred to [34, 35]. However, they can not output better predictions than Eq. (9)~(12).

Substituting Eqs. (9)~(11) into Eq. (4) and setting $Z_c^b = 6$ establish an equation regarding $\phi_{RCP}^b$, $X_s$ and $R_r$, and thus we can solve $\phi_{RCP}^b$ on the $R_r$-$X_s$ space. To better demonstrate the tendency of $\phi_{RCP}^b$ with $R_r$ and $X_s$, and further emphasize the comparison between $\phi_{RCP}^b$ and $\phi_{RCP}^m$, we introduce a relative value $\Delta\bar{\phi}_{RCP}^b$ to describe the difference as,

$$\Delta\bar{\phi}_{RCP}^b = \frac{\phi_{RCP}^b - \phi_{RCP}^m}{\phi_{RCP}^m}. \tag{13}$$

We firstly solve the distribution of $\phi_{RCP}^b$ based on the original expression in [33], i.e., $n_l = 1$ and $n_s = 1$, and compare our prediction against the reported data in [25]. As shown in Figure 1(a), the prediction of our theory can capture the tendency of $\phi_{RCP}^b$, i.e., an asymmetric distribution along $X_s$, which is also in agreement with direct observations from other experiments and simulations [12, 16, 36]. Furthermore, the solution quantitatively matches the reported data in the region $R_r \to 1$, though an obvious deviation is observed in the moderate-$R_r$ region ($R_r < 0.5$). This drawback is likely to be conquered by trying other statistical models, such as Percus-Yevick theory [37] and the Carnahan-Starling equation [31]. However, we stick to the current scheme since its solution is accurate enough for the region of our interest. More important, the obtained theoretical results predict a valley region ($\Delta\bar{\phi}_{RCP}^b < 0$) near the boundaries ($R_r = 1$ and $X_s = 0$), as highlighted by the black dash lines in Figure 1(b), constituting the so-called loose jamming state. To further confirm the existence of this valley region, numerical evidence is provided below.

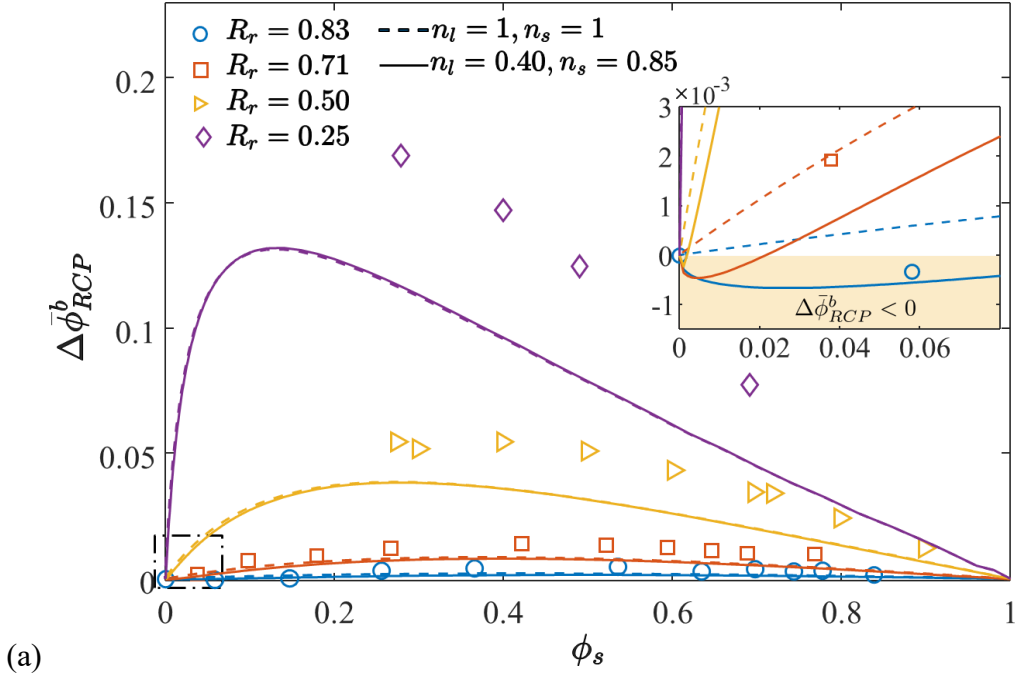

(a)

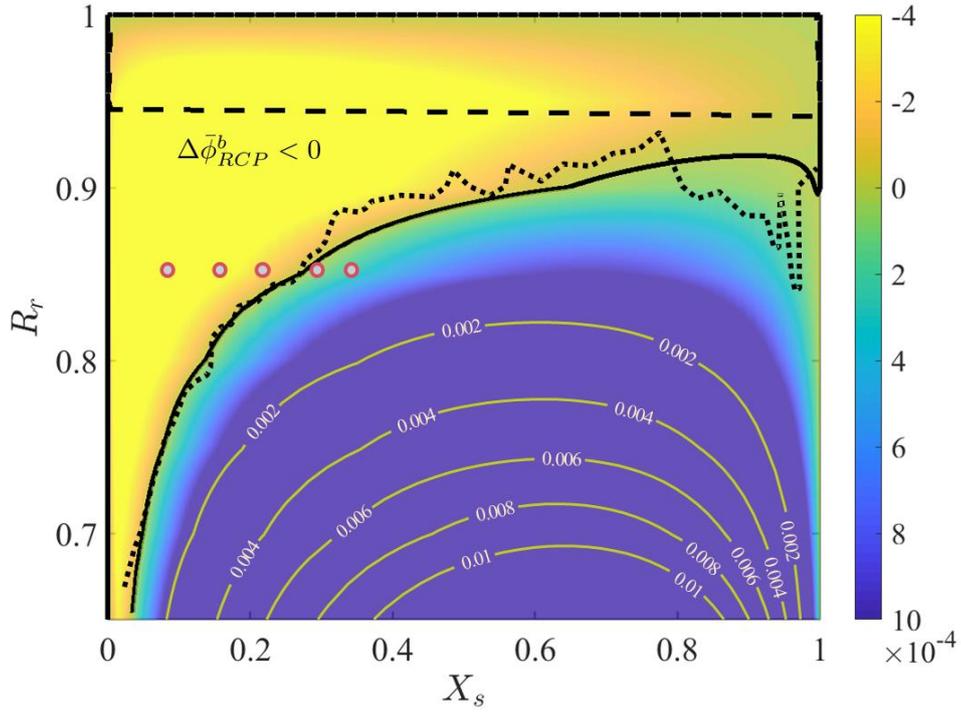

(b)

Figure 1. (a) Relative packing factor $\Delta\bar{\phi}^b_{RCP}$ vs. volume fraction of the small component, $\phi_s = X_s R_r^3 / (X_s R_r^3 + X_l)$. Open symbols are data reported in [25]; dash lines are theoretical prediction base on the Eqs. (9)~(11) with $n_l = 1$ and $n_s = 1$ while solid lines are the one with $n_l = 0.40$ and $n_s = 0.85$. The subfigure is a zoom-in view of near-boundary region ($X_s \to 0$). (b) The contour plot of the relative packing factor $\Delta\bar{\phi}^b_{RCP}$ from simulations on the $R_r$-$X_s$ space. The valley region appears in yellow. The dashed line represents the prediction by the original expression of $g_{ij}(d_{ij})$ in [33], with $n_l = 1$ and $n_s = 1$; the dotted line is extracted from simulation results; the solid line is predicted by Eqs. (9)~(11) with $n_l = 0.40$ and $n_s = 0.85$, selected for fitting the simulation data. Symbol ○ can be referred to Figure 2 for further statistical analyses of the local packing structure of spheres.

*Simulations*

Numerical simulations are performed by an inhouse code, KIT-DEM [18, 38]. Each configuration is identified by a unique size ratio $R_r$ and mole fraction $X_s$, and is repeated for five samples. Each packing contains 5000 frictionless non-cohesive spherical particles which are randomly packed within a periodic cubic domain as an initial state [39]. The Hertzian contact model is adopted and the elasticity modulus is set large enough to guarantee the final particle-particle overlap is smaller than 0.1% particle radius so that the hard-sphere assumption can be satisfied. All packings are uniaxially compressed using a quasi-static protocol [38] until reaching the jamming point where the average coordination number $Z_c$ increases to six and the increment rate of $Z_c$ starts to demonstrate a sharp decay [12]. The critical packing factor is then obtained at the jamming point, and for each configuration $\phi_{RCP}$ is calculated as the mean value over its five repeated cases, and the standard deviation at the valley is $1.3 \times 10^{-4}$ while the corresponding $\Delta \bar{\phi}_{RCP}^b = -1.7 \times 10^{-3}$.

*Results and Discussion*

The relative packing factor $\Delta \bar{\phi}_{RCP}^b$ with respect to $R_r$ and $X_s$ extracted from the conducted simulations is plotted in Figure 1(b). The introduced indexes $n_l$ and $n_s$ in Eqs. (9) and (10) are adjusted to ensure that the theoretical solution matches simulation results, and a good agreement can be obtained when $n_s = 0.40$ and $n_l = 0.85$, as shown in Figure 1(b) where the original expression in [24] (with $n_l = 1$ and $n_s = 1$), is also included to compare with the simulation results. To be safe, we also test this modified version against the original version and the reported data. In Figure 1(a), solutions based on the different groups of $n_l$ and $n_s$ are almost the same over the most range of $\phi_s$, except in the range of $\phi_s \to 0$, see the subfigure. This suggests that modifying the additional terms in Eqs. (9) and (10) can only effectively influence the near-boundary range rather than the moderate $R_r$-$X_s$ space. Moreover, the proposed expression with modified $n_l$ and $n_s$ is capable of predicting the loose jamming region, which is not only confirmed by our simulation results but also supported by the data reported in [25] on a qualitative basis.

In addition to the statistical view on the binary packing and its loose jamming state, we further provide a meso-scale insight into this unexplored valley with the help of the pair correlation function $p(r)$ [40, 41], which can measure the probability density of possible distances between centres of two particles ($\vec{c}_i$ and $\vec{c}_j$) within a radial space, and thus it is defined as

$$p(r) = \frac{1}{\rho(N_s+N_l)} \sum_{\vec{c}_i} \sum_{\vec{c}_j} \delta[(|\vec{c}_i - \vec{c}_j| - r) \cdot (r + b - |\vec{c}_i - \vec{c}_j|)] \frac{1}{4\pi r^2 b}. \quad (14)$$

The peak value of $p(r)$ occurs corresponding to a certain particle-particle configuration, which is classified into three neighbouring layers (not limited to three, but the most related ones are discussed here). Figure 2 (a)~(e) demonstrate a transition from the loose to normal jamming state. From the variation of $p(r)$ with $X_s$, the third peak moves leftwards from $2.65d_l$, where large particles dominate the third layer, to smaller than $2.56d_l$, suggesting small particles break through the third layer into the second. In other words, the jamming state turns to normal (with the third peak of $2.65d_l$ in $p(r)$) from the loose one (with the third peak of $2.56d_l$ in $p(r)$) as we increase the amount of small spheres in the mixture, illustrated in Figure 2(f). Additionally, the second peak becomes obscure even vanishes after the jamming state transition, see (d) and (e), indicating small particles contribute more to the second and third peaks in $p(r)$, and even dominating the third layer. Therefore, the impact of small particles on the third layer can be regarded as a meso-scale judgement on the jamming state, and the critical position of third peak on $p(r)$ defines the phase boundary.

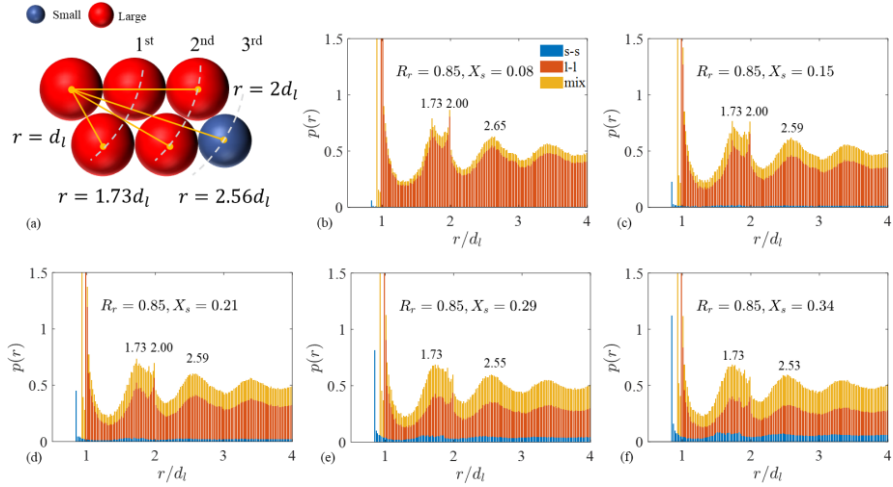

Figure 2. The variation of pair correlation function $p(r)$ of $R_r = 0.85$ crossing the border of loose jamming state with increasing $X_s$, i.e., from (b-d) locating within the valley region to (e-f) locating outside, and the corresponding positions on the $R_r$-$X_s$ space can be referred to the symbol ⊙ in Figure 1(b). The values of $r/d_l$ at the local peaks of $p(r)$ are indicated, and can be referred to the particle-particle topology in (a).

*Conclusions*

In this letter, we have developed a theoretical model by extending the recent Zaccone's scheme [23] for predicting the RCP of binary granular assemblies. The proposed model has been firstly validated against the data reported in [25] focusing on the relationship between the packing factor and volume fraction of small spheres. The analytical solution predicts an unexpected valley region on $R_r$-$X_s$ space, i.e., the loose jamming state, at which $\phi_{RCP}^b$ is smaller than $\phi_{RCP}^m$ against the common observation within the moderate $R_r$-$X_s$ range that $\phi_{RCP}^b$ should be always larger than $\phi_{RCP}^m$. Subsequently, this valley indicating the phase transition from loose to normal jamming states has been further confirmed through DEM simulations and the exact phase boundary of the loose jamming state is established. The simulation data also provided a meso-scale insight into this observed transition with increasing fraction of smaller particles, during which small particles gradually dominate the third layer and are mixed into the second layer. Finally, we linked the statistical description of the loose jamming state with a meso-scale explanation based on particle-particle topology.

In the future, this theoretical approach can be possibly extended to polydisperse granular media. Compared with the binary packing, it seems likely to realize much looser jamming packings by adjusting size distribution. Since mechanical and transport properties of granular materials are related to packing structures, this adjustable jamming state may lead to important engineering practices, such as lightweight concrete design [42], microstructure design of battery electrodes[43], and optimized granular beds for water retention [44, 45], filtration [46, 47], and thermal energy storage [48, 49].

This work was supported by State Key Laboratory for Strength and Vibration of Mechanical Structures under SV2021-KF-27.